\documentclass[aps,prl,twocolumn,showpacs]{revtex4}
\usepackage{amsmath}
\usepackage{amssymb}
\usepackage{graphicx}
\usepackage{dcolumn}
\usepackage{bm}

 \draft

\begin{document}

\title{Van der Waals forces and spatial dispersion}
\author{L.~P.~Pitaevskii}
\date{\today}

\begin{abstract}
A version of the Green's functions theory of the Van der Waals forces which
can be conveniently used in the presence of spatial dispersion is presented.
The theory is based on the fluctuation-dissipation theorem and is valid for
interacting bodies, separated by vacuum. Objections against theories
acounting for the spatial dispersion are discussed.
\end{abstract}

\affiliation{CNR INFM-BEC and Department of Physics, University of Trento, I-38050 Povo,
Trento, Italy; \\
Kapitza Institute for Physical Problems, Kosygina 2, 119334 Moscow, Russia}
\maketitle

\textbf{Introduction.} It was recently discovered that effects of the
spatial dispersion are quite important for calculations of the Van der Waals
forces\cite{VdW} between conducting bodies at finite temperatures (see \cite%
{Sernelius,Esquivel,Pitaevskii,Lam,Svetovoy}). On the other hand, an opinion
has been expressed in several papers that the theory of the forces in the
present form is not valid in the presence of this dispersion (see \cite%
{Most1,KMPRA07,Most2}). The goal of this paper is to prove rigorously that the theory
of the Van der Waals forces, as it was formulated in the terms of Green's
functions in papers \cite{DP,DLP} (DLP below, see also \cite{LP9}), can be
used for calculation of the forces for bodies \textit{separated by vacuum},
even if the spatial dispersion is important for electromagnetic properties
of the bodies.

To my understanding, the doubts about the validity of the theory in the
presence of the spatial dispersion is based on a remark in a paper by Barash
and Ginzburg \cite{BG75}. In this paper, the contribution to the free energy
of the long-wave fluctuations of the electromagnetic field is presented in
the form%
\begin{eqnarray}
\Delta F &=&-k_{B}T\sum_{n=0}^{\infty }\;^{\prime }\mathrm{Sp}\ln \left[ 
\mathcal{D}_{n}\left( \mathcal{D}_{n}^{(0)}\right) ^{-1}\right]  \label{BG}
\nonumber \\
&&+k_{B}T\sum_{n=0}^{\infty }\;^{\prime }\int_{0}^{1}\zeta ^{2}d\zeta 
\mathrm{Sp}\left[ \frac{\partial \Pi _{n}\left( \zeta \right) }{\partial
\zeta }\mathcal{D}_{n}\left( \zeta \right) \right]  \ ,
\end{eqnarray}%
where $\mathcal{D}_{n}$ is the Matsubara Green's function of the system
under considerations and $\mathcal{D}_{n}\left( \zeta \right) $ \ and $\Pi
_{n}\left( \zeta \right) $ are correspondingly the Green's function and the
polarization operator of an auxiliary system in which interaction with the
long-wave field is defined by the "charge" $\sqrt{\zeta }e$ instead the
actual charge $e$ (see \cite{BG75}, Eq. (54)). The prime\ sign means that
the $n=0$ term is taken with the coefficient $1/2$. The first term in (\ref%
{BG}) corresponds to the DLP theory, while the second gives an additional
contribution. The authors noticed that in the presence of significant
spatial dispersion the second term cannot be neglected and consequently the
free energy cannot be expressed only in terms of the Green's functions of
the actual system. 

However, in my opinion this statement is related to the
general case, when bodies are embedded in a liquid with spatial dispersion.
Results of the DLP theory are indeed difficult to generalize to this case
due to the non-local nature of the stress tensor in a such liquid. In the
important case when the interacting bodies are separated by vacuum this difficulty
does not arise. Indeed, forces acting between any bodies separated \textit{by vacuum} 
can be calculated by averaging of the vacuum Maxwell stress tensor. 
Corresponding quadratic combinations of the field strengths can in turn 
be expressed in terms of the retarded Green's function of the field using 
the \textit{exact}  fluctuation-dissipation theorem (see \cite{LP9}). 
This Green's function, of course, must be calculated 
taking into account the spatial dispersion in the bodies, when it is important. 
This programme wil be realized below.
It is convenient to use an approach which is
different both from \cite{DP,DLP} and \cite{BG75}. (The basic idea of the
derivation was expressed in short in \cite{Pitaevskii}.)

\textbf{Maxwell stress tensor.} The force acting in vacuum on a body due to
electromagnetic fluctuations is equal to\cite{foot1}%
\begin{equation}
f_{i}=-\oint \sigma _{ik}dS_{k}\ ,  \label{force}
\end{equation}%
where the integration is performed over any surface enclosing only the given
body and $\sigma _{ik}$ is the average value of the Maxwell stress tensor of
electromagnetic fluctuations in vacuum:%
\begin{equation}
\sigma _{ik}\left( \mathbf{r}\right) =-\frac{\left\langle E^{2}\right\rangle 
}{8\pi }\delta _{ik}+\frac{\left\langle E_{i}E_{k}\right\rangle }{4\pi }-%
\frac{\left\langle B^{2}\right\rangle }{8\pi }\delta _{ik}+\frac{%
\left\langle B_{i}B_{k}\right\rangle }{4\pi }\ .  \label{Maxwell}
\end{equation}%
Here we introduced notations for the averages from operator expressions:%
\begin{eqnarray}
\left\langle E_{i}E_{k}\right\rangle &=&\left\langle \hat{E}_{i}(\mathbf{r}%
,t)\hat{E}_{k}(\mathbf{r},t)+\hat{E}_{k}(\mathbf{r},t)\hat{E}_{i}(\mathbf{r}%
,t)\right\rangle /2\ ,  \label{EE} \\
\left\langle E^{2}\right\rangle &=&\left\langle E_{i}E_{i}\right\rangle 
\notag
\end{eqnarray}%
\ and \ 
\begin{eqnarray}
\left\langle B_{i}B_{k}\right\rangle &=&\left\langle \hat{B}_{i}(\mathbf{r}%
,t)\hat{B}_{k}(\mathbf{r},t)+\hat{B}_{k}(\mathbf{r},t)\hat{B}_{i}(\mathbf{r}%
,t)\right\rangle /2\ ,  \label{BB} \\
\left\langle B^{2}\right\rangle &=&\left\langle B_{i}B_{i}\right\rangle \ . 
\notag
\end{eqnarray}%
I consider only the equilibrium situation and averaging is taken using the
Gibbs statistics. Thus the problem reduces to calculation of the average of
quadratic combinations of field strength operators. Notice, that this method
of calculation of forces can be applied only in vacuum, because there is in
general no equation for the stress tensor of non-static fields in an
arbitrary medium.

\textbf{Fluctuation-dissipation theorem.} To define these averages one can
use the exact fluctuation-dissipation theorem (FDT). To apply this
theorem one must consider correlation functions of the fields in different
points in space and different moments of time:%
\begin{eqnarray}
&&S_{ik}^{E}\left( \mathbf{r},\mathbf{r}^{\prime };\tau \right)   \label{SE}
\\
&=&\left\langle \hat{E}_{i}(\mathbf{r},t+\tau )\hat{E}_{k}(\mathbf{r}%
^{\prime },t)+\hat{E}_{k}(\mathbf{r}^{\prime },t)\hat{E}_{i}(\mathbf{r}%
,t+\tau )\right\rangle /2\ \ ,  \notag
\end{eqnarray}%
where we took into account the uniformity with respect to time. The magnetic
function $S_{ik}^{B}\left( \mathbf{r},\mathbf{r}^{\prime };\tau \right) $ is
defined in an analogous way. We need also the correlation function of the
components of the 4-vector potential. Following \cite{DP} we will use the
gauge with the scalar potential $\varphi =0.$ Then 
\begin{equation}
\mathbf{\hat{E}=-}\frac{1}{c}\frac{\partial \mathbf{\hat{A}}}{\partial t}\ ,%
\mathbf{\hat{B}=\mathrm{{curl}\ \hat{A}\ }}  \label{AEB}
\end{equation}%
and $S_{ik}^{E}$ , $S_{ik}^{B}$ \ can be expressed as (we put $c=\hbar =1$
below)%
\begin{eqnarray}
S_{ik}^{E}\left( \mathbf{r},\mathbf{r}^{\prime };\tau \right)  &=&\frac{%
\partial ^{2}}{\partial t^{2}}S_{ik}^{A}\left( \mathbf{r},\mathbf{r}^{\prime
};\tau \right) ,  \label{SEA} \\
S_{ik}^{B}\left( \mathbf{r},\mathbf{r}^{\prime };\tau \right)  &=&\left( 
\mathrm{curl}\right) _{is}\left( \mathrm{{curl}^{\prime }}\right)
_{kl}S_{sl}^{A}\left( \mathbf{r},\mathbf{r}^{\prime };\tau \right)   \notag
\end{eqnarray}%
in the terms of the correlation function of the components of $\mathbf{\hat{A%
}}$:%
\begin{eqnarray}
&&S_{ik}^{A}\left( \mathbf{r},\mathbf{r}^{\prime };\tau \right)   \label{SA}
\\
&=&\left\langle \hat{A}_{i}(\mathbf{r},t+\tau )\hat{A}_{k}(\mathbf{r}%
^{\prime },t)+\hat{A}_{k}(\mathbf{r}^{\prime },t)\hat{A}_{i}(\mathbf{r}%
,t+\tau )\right\rangle /2\ \ .  \notag
\end{eqnarray}%
Expand now the function $S_{ik}^{A}$ in the Fourier integral with respect to
time:%
\begin{equation}
S_{ik}^{A}\left( \mathbf{r},\mathbf{r}^{\prime };\tau \right) =\int_{-\infty
}^{\infty }e^{-i\omega t}S_{ik}^{A}\left( \mathbf{r},\mathbf{r}^{\prime
};\omega \right) \frac{d\omega }{2\pi }\ .  \label{So}
\end{equation}%
The central point of the proof is that the fluctuation-dissipation theorem
allows us to express $S_{ik}^{A}\left( \mathbf{r},\mathbf{r}^{\prime
};\omega \right) $ in the terms of the retarded Green's function of the
electromagnetic field. Indeed, the interaction of the field with the current
is given by the equation%
\begin{equation}
V=-\int \mathbf{\hat{A}}(\mathbf{r})\mathbf{\hat{\jmath}}\left( \mathbf{r}%
\right) d\mathbf{r\ .}  \label{int}
\end{equation}%
Let us consider the vector-potential which is induced by a classical
external current with the density $\mathbf{j}^{ext}\left( \mathbf{r}%
,t\right) =\left[ \mathbf{j}_{\omega }^{ext}\left( \mathbf{r}\right)
e^{-i\omega t}+\mathbf{j}_{-\omega }^{ext}\left( \mathbf{r}\right)
e^{i\omega t}\right] /2.$ In the linear approximation the induced potential
can be expressed in the terms of the retarded Green's function of the
vector-potential:%
\begin{equation}
A_{i\omega }^{ind}\left( \mathbf{r}\right) =-\int D_{ik}^{R}\left( \mathbf{%
r,r}^{\prime };\omega \right) j_{k\omega }^{ext}\left( \mathbf{r}^{\prime
}\right) d\mathbf{r}^{\prime }\ ,  \label{ext}
\end{equation}%
where, according the Kubo equation, 
\begin{eqnarray}
&&D_{ik}^{R}\left( \mathbf{r,r}^{\prime };\omega \right)   \label{Kubo} \\
&=&-i\int_{0}^{\infty }e^{i\omega t}\left\langle \hat{A}_{i}(\mathbf{r},\tau
)\hat{A}_{k}(\mathbf{r}^{\prime },0)+\hat{A}_{k}(\mathbf{r}^{\prime },0)\hat{%
A}_{i}(\mathbf{r},\tau )\right\rangle d\tau \ .  \notag
\end{eqnarray}%
Then according to the fluctuation-dissipation theorem (see \cite{LL5},
sections 122-125 and \cite{LP9}, sections\ 75-76),%
\begin{eqnarray}
&&S_{ik}^{A}\left( \mathbf{r},\mathbf{r}^{\prime };\omega \right) 
\label{FDT} \\
&=&\frac{1}{2}i\coth \left( \frac{\hbar \omega }{2k_{B}T}\right) \left\{
D_{ik}^{R}\left( \mathbf{r,r}^{\prime };\omega \right) -\left[
D_{ki}^{R}\left( \mathbf{r}^{\prime }\mathbf{,r};\omega \right) \right]
^{\ast }\right\} \ .  \notag
\end{eqnarray}%
If the bodies are not magnetoactive, the $D$-function satisfies the Onsager
symmetry relation $D_{ik}^{R}\left( \mathbf{r,r}^{\prime };\omega \right)
=D_{ki}^{R}\left( \mathbf{r}^{\prime }\mathbf{,r};\omega \right) .$ Then we
find finally%
\begin{equation}
S_{ik}^{A}\left( \mathbf{r},\mathbf{r}^{\prime };\omega \right) =-\coth
\left( \frac{\hbar \omega }{2k_{B}T}\right) \mathrm{Im}D_{ik}^{R}\left( 
\mathbf{r,r}^{\prime };\omega \right) \ .  \label{FDT2}
\end{equation}%
This equation together with (\ref{SEA}) solves the problem of calculating
electromagnetic fluctuations in the terms of $D^{R}.$ One has%
\begin{equation}
\left\langle E_{i}E_{k}\right\rangle =\mathrm{Im}\int_{-\infty }^{\infty
}\omega ^{2}\coth \left( \frac{\hbar \omega }{2k_{B}T}\right)
D_{ik}^{R}\left( \mathbf{r,r};\omega \right) \frac{d\omega }{2\pi }
\label{EiEkD}
\end{equation}%
and%
\begin{eqnarray}
\left\langle B_{i}B_{k}\right\rangle  &=&-\mathrm{Im}\int_{-\infty
}^{\infty }\coth \left( \frac{\hbar \omega }{2k_{B}T}\right)   \label{BiBkD}
\\
&&\times \left[ \left( \mathrm{curl}\right) _{is}\left( \mathrm{curl}%
^{\prime }\right) _{kl}D_{sl}^{R}\left( \mathbf{r,r}_{1};\omega \right ) %
\right] _{\mathbf{r}=\mathbf{r}_{1}}\frac{d\omega }{2\pi }\ .  \notag
\end{eqnarray}

Equation (\ref{Kubo}) is not very useful for calculating the function $%
D_{ik}^{R}$. However, $D^{R}$ can be calculated directly using equation (\ref%
{ext}). Indeed, let us place a unit point-like current directed in the $l$%
-direction at the point $\mathbf{r}^{\prime }=\mathbf{r}_{0},$ i. e. put 
\begin{equation}
j_{k\omega }^{ext}\left( \mathbf{r}^{\prime }\right) =-\delta _{il}\delta
\left( \mathbf{r}^{\prime }-\mathbf{r}_{0}\right) \ .  \label{source}
\end{equation}%
Using a proper theory of electrodynamic properties of the interacting
bodies, one must calculate the linear response $A_{il\omega }^{ind}\left( 
\mathbf{r,r}_{0}\right) $ for the source (\ref{source}). Then according to (%
\ref{ext}) we get%
\begin{equation}
D_{ik}^{R}\left( \mathbf{r,r}_{0};\omega \right) =A_{ik\omega }^{ind}\left( 
\mathbf{r,r}_{0}\right) \ .  \label{Dsource}
\end{equation}%
It is important that it is enough for the calculation of the tensor to
consider the source placed in vacuum between the bodies. This, for example,
allows in a simple plain geometry to express $D_{ik}^{R}$ in the terms of
the reflection amplitude on the surfaces of the bodies. Expression (\ref%
{Dsource}) for $D_{ik}^{R}$ diverges at $\mathbf{r\rightarrow r}_{0}.$ This
divergency can be eliminated by subtracting the response $\bar{A}_{il\omega
}^{ind}\left( \mathbf{r,r}_{0}\right) $ for the source (\ref{source}),
calculated for free space.

Notice that according to (\ref{EiEkD})-(\ref{BiBkD}) \ the stress tensor can
be presented in the form%
\begin{equation}
\sigma _{ik}=\mathrm{Im}\int_{0}^{\infty }\coth \left( \frac{\hbar \omega }{%
2k_{B}T}\right) \sigma _{ik}^{R}\left( \omega \right) \frac{d\omega }{2\pi }
\label{sigmaR}
\end{equation}%
where $\sigma _{ik}^{R}\left( \omega \right) $ is an analytical function
having no singularities in the upper half-plane $\omega .$ Displacing the
contour of integration to the imaginary axes of $\omega $, as in the
original Lifshitz paper\cite{Lif}, and taking the residues in the poles of
the $\coth $, we reduce (\ref{sigmaR}) to the Matsubara form:%
\begin{equation}
\sigma _{ik}=\frac{k_{B}T}{\hbar }\sum_{n=0}^{\infty }\;^{\prime }\sigma
_{ik}^{R}\left( i\xi _{n}\right) ,\xi _{n}=2\pi k_{B}T/\hbar \ .
\label{Matsubara}
\end{equation}

\textbf{Conclusion.} The general theory which I present here does not give,
of course, a magic prescription for solution of all problems of Van der
Waals forces in the presence of spatial dispersion. Calculations of the $%
D^{R}$ function as a response to a point source can be difficult and in any
case demands a sort of microscopic theory. However, I believe that reducing
the problem to the exact fluctuation-dissipation theorem gives a solid
foundation for different theoretical approaches.

It must be noted however that in a recent paper \cite{Most2} the authors
express doubts about the validity of the Lifshitz theory based on this
theorem for materials with finite d.c. conductivity. Because FDT is a 
\textit{rigorous} theorem of quantum statistical mechanics, such doubts must
have a very serious foundation. On the contrary, the consideration of the
authors are, in my opinion, quite superficial. One can read in \cite{Most2}
(I omit references):

\textquotedblleft Lifshitz derived his famous formulae under the condition
of thermal equilibrium. This means that not only $T=$ const, but also all
irreversible processes connected with the dissipation of energy into heat
have already been terminated ... The Drude-like dielectric function ... is
derived from the Maxwell equations with a real drift current of conduction
electrons $\mathbf{j}=\sigma _{0}\mathbf{E}$ initiated by the external
electric field $\mathbf{E}$ ... The drift current is an irreversible process
which brings a system out of thermal equilibrium. ... The real current leads
to Joule's heating of the Casimir plates (Ohmic losses) ... To preserve the
temperature constant, one should admit that there exists an unidirectional
flux of heat from the medium to the heat reservoir ... Such interactions
between a system and a heat reservoir are prohibited by the definition of
thermal equilibrium. Although the screening and diffusion effects really
occur in an external electric field, they are also related to physical
situations out of thermal equilibrium. The reason is that the diffusion
current is determined by a nonzero gradient of charge carrier density,
whereas for homogeneous systems in thermal equilibrium the charge carrier
density must be homogeneous.\textquotedblright

I believe that these considerations are wrong. Of course, one can say that
the fluctuating electric field heats a body. However, in equilibrium this
heating is compensated by emission of radiation by the body. This exact
compensation is ensured by the detailed balance principle. The sentence
about termination of dissipation of energy is particularly odd in relation
to the fluctuation-dissipation theorem, which just connects the energy
dissipation with fluctuations. It is not clear, by the way, why the authors
worry only about the Ohmic losses. At finite frequencies all real materials
dissipate energy.

The statement that the screening and diffusion effects are related to
physical situations out of thermal equilibrium is also wrong. It is
well-known that the Boltzmann distribution in the electric field, which was
used in \cite{Pitaevskii} for describing screening, is an equilibrium
distribution. Actually, in equilibrium the diffusion current is compensated
by the mobility of carriers due to an electric field.

The authors of \cite{Most2} claim also that \textquotedblleft for
dielectrics whose charge carrier density is temperature-independent (for
such materials conductivity goes to zero with $T$ not due to the vanishing $%
n $ but due to the vanishing mobility of the charge carriers) the
generalization of the Lifshitz theory taking into account the screening
effects is shown to violate the Nernst theorem\textquotedblright\ and is
consequently \textquotedblleft in contradiction with
thermodynamics\textquotedblright . I believe that this statement is a
result of a pure misunderstanding. The materials under discussion are
amorphous glass-like \textit{disordered} bodies. Conductivity goes to zero
with $T$ in such materials due to localization of the charge carriers just
because of the disorder. The point is that the Nernst theorem \textit{is not
valid} for these disordered bodies. It is well known that they have a big finite entropy at zero
temperature. Localized carriers also contribute to this residual entropy and
the calculation of a small correction to its value due to the Van der Waals
interaction scarcely has a physical meaning. Of course, the existence of
disordered bodies at $T=0$ itself does not contradict statistical mechanics. They
are simply not at an equilibrium state at low temperatures due to a very
long relaxation time. However, the role for our phenomena of the long
relaxation time in these bodies is not clear at present and is worth of
careful investigation.


I thank Yu. Barash, V. Svetovoy and E. Taylor for useful discussions.


\begin{thebibliography}{99}
\bibitem{VdW} I use the generic term \textquotedblleft Van der Waals
forces\textquotedblright\ for long-range forces between neutral objects in
any conditions. Thus I do not distinguish between London, Casimir,
Casimir-Polder and Lifshitz forces.

\bibitem{Sernelius} B.~E.~Sernelius, J. Phys. A: Math. Gen. \textbf{39},
6741 (2006).

\bibitem{Esquivel} R~Esquivel-Sirvent et. al., J. Phys. A: Math. Gen. 
\textbf{39}, 6323 (2006).

\bibitem{Pitaevskii} L. P. Pitaevskii, Phys. Rev. Lett. \textbf{101},
163202 (2008).

\bibitem{Lam} D. A. R. Dalvit and S. K. Lamoreaux, Phys. Rev. Lett. \textbf{101},
163203 (2008).

\bibitem{Svetovoy} V. B. Svetovoy, Phys. Rev. Lett. \textbf{101},
163603 (2008).



\bibitem{Most1} V. M. Mostepanenko et al., J. Phys. A: Math. Gen. \textbf{39}%
, 6589 (2006).

\bibitem{KMPRA07} G. L. Klimchitskaya and V. M. Mostepanenko, Phys. Rev. B 
\textbf{75}, 036101 (2006).

\bibitem{Most2} G.~L.~Klimchitskaya, U.~Mohideen, and V.~M.~Mostepanenko,
ArXiv:0802.2698v3., J. Phys. A: .Math. Theor. \textbf{41}, 432001 (2008).

\bibitem{DP} I.~E.~Dzyaloshinskii and L.~P.~Pitaevskii, Sov. Phys. JETP 
\textbf{9}, 1282 (1959).

\bibitem{DLP} I.~E.~Dzyaloshinskii, E. M. Lifshitz, and L.~P.~Pitaevskii,
Adv. Phys. \textbf{10}, 165 (1961).

\bibitem{LP9} E.~M.~Lifshitz and L.~P.~Pitaevskii, \emph{Statistical
Physics, Part 2}, Pergamon Press, Oxford, 1984.

\bibitem{BG75} Yu. S. Barash and V. L. Ginzburg, Sov. Phys. Usp. \textbf{18}%
, 305 (1975).

\bibitem{foot1} There is a statement in the paper \cite{Welsch} that it is
\textquotedblleft not possible to correctly calculate the force on one of
two or more bodies in a medium from a stress tensor whose divergence yields
the correct force density only within the medium\textquotedblright . This
statement contradicts to the meaning of the stress tensor and is obviously
wrong. The integral (\ref{force}) gives the full momentum recieved by the
body per unit of time, i. e. the force.

\bibitem{Welsch} C. Raabe and D.-G. Welsch, Phys. Rev. A \textbf{73}, 047802
(2006).


\bibitem{LL5} L.~D. Landau and E.~M.~Lifshitz, \emph{Statistical Physics,
Part 1, }Pergamon Press, Oxford, 1980.



\bibitem{Lif} E.~M.~Lifshitz, Sov. Phys. JETP \textbf{2}, 73 (1956). 

\end{thebibliography}
\end{document}